\documentclass[useAMS,usenatbib]{mn2e}
\usepackage[dvips]{graphicx}
\usepackage{amsmath}
\usepackage{amsfonts}
\usepackage{amssymb}

\title[Transit timing effects due to an exomoon II]{Transit timing effects due to an exomoon II}
\author[David M. Kipping]{David M. Kipping$^{1}$\thanks{E-mail:
d.kipping@ucl.ac.uk}\footnotemark[1]\\
$^{1}$Department of Physics and Astronomy, University College London, \\
       Gower Street, London WC1E 6BT, UK}
\begin{document}

\date{Accepted 2009 April 3. Received 2009 March 12; in original form 2009 January 12}

\pagerange{\pageref{firstpage}--\pageref{lastpage}} \pubyear{2008}

\maketitle

\label{firstpage}

\begin{abstract}
In our previous paper, we evaluated the transit duration variation (TDV) effect for a co-aligned planet-moon system at an orbital inclination of $i=90^{\circ}$.  Here, we will consider the effect for the more general case of $i \leq 90^{\circ}$ and an exomoon inclined from the planet-star plane by Euler rotation angles $\alpha, \beta$ and $\gamma$.  We find that the TDV signal has two major components, one due to the velocity variation effect described in our first paper and one new component due to transit impact parameter variation.  By evaluating the dominant terms, we find the two effects are additive for prograde exomoon orbits, and deductive for retrograde orbits.  This asymmetry could allow for future determination of the orbital sense of motion.  We re-evaluate the ratio of TDV and TTV effects, $\eta$, in the more general case of an inclined planetary orbit with a circular orbiting moon and find that it is still possible to directly determine the moon's orbital separation from just the ratio of the two amplitudes, as first proposed in our previous paper.
\end{abstract}


\section{Introduction}

With exoplanet detection rates soaring, it is now becoming increasingly possible to characterise these alien worlds.  Part of this characterisation will undoubtedly involve determining if exoplanets have their own moons, so called exomoons.  The theoretical foundations of exomoon detection using transits were first laid down by \citet{sar99}, \citet{dee02}, \citet{sza06} and \citet{sim07}.  In these papers, the emerging theme of using transit timing variation (TTV) as a detection tool was advocated.  In our previous paper, \citet{kip09} (hereafter K09), we showed that an exomoon should induce not only a transit time variation (TTV) effect but also a transit duration variation (TDV) effect on the host planet.  The two effects were predicted to exhibit a $\pi/2$ phase difference which could be used as the hallmark signature of an exomoon.

In our previous work, an underlying assumption was coplanarity within the system.  We assumed that both the planet's orbital inclination angle, $i$, was $90^{\circ}$ and that the moon's orbit was completely coplanar with the planet-star orbit.  In this paper, we will extend the theoretical framework to include non-coplanarity.  As a result of this consideration, we predict that the TDV effect due to an exomoon has infact two primary constituents: i) a velocity (V) component ii) a transit impact parameter (TIP) component.

The V-component is the same effect we described in our previous work, where the velocity of the planet is perturbed by the moon's presence.  The TIP-component is a new effect which is due to the planet moving between higher and lower impact parameters as a result of the wobbling.  Since transit duration is a strong function of impact parameter, then even slight changes can induce a TDV effect.

This additional TDV component acts constructively with the V-component in the case of a prograde exomoon orbit and destructively for a retrograde orbit.  With most large moons taking prograde orbits within the solar system, it would seem reasonable to expect constructive interference to be the typical case.  As a result, the expected TDV signal from an exomoon is even more detectable.  Furthermore, we predict this asymmetry could allow for a determination of the satellite's sense of orbital motion.

\section{The TIP-Component of the TDV Effect}

In general, we posit that there exists two dominant components of the TDV effect.  The V-component is caused by an exomoon perturbing the planet's velocity as it orbits the host star.  This effect was discussed in depth in our previous paper and details can be found in K09.  The second effect is the one we will concentrate on in this work and we label it as the transit impact parameter (TIP) component.

Consider a side-on view of a planetary transit as shown in figure 1.  As demonstrated by \citet{sea03}, the transit duration is a strong function of the impact parameter of the transit, $b$, which is given by:

\begin{equation}
b = \frac{r_P \cos i}{R_*} = \frac{q}{R_*}  
\end{equation}

The variable $q$ is shown in figure 1 and denotes the distance between the observer's line-of-sight to the planet's centre, during the transit.  It is this distance, $q$, which the transit duration is particularly sensitive to.  If $q \rightarrow 0$, the planet transits across the star's equator, which is the star's widest point, thus giving a very long transit duration.  If $q \rightarrow R_*$, then the planet only grazes the star during the transit event and so we expect a very short duration.

\begin{figure}
\begin{center}
\includegraphics[width=8.4 cm]{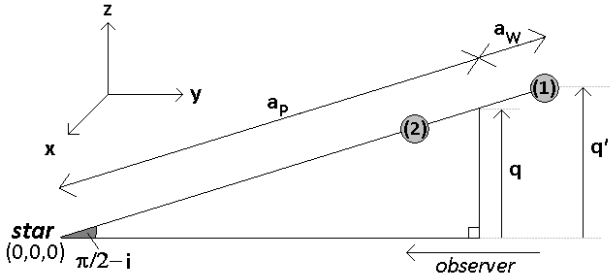}
\caption{\emph{Cartoon of the side-on view of the star-planet-moon system.  In this schematic, the star lies at the apex of the lines in the bottom left, the observer lies at $y = +\infty$ and the exomoon is not shown.  The wobble of the planet is represented by the two grey spheres, (1) and (2), being the planet's maximal positions.  The presence of a moon causes distance a perturbation in the distance $q$.}} \label{fig:fig1}
\end{center}
\end{figure}

Now consider adding an exomoon.  As discussed in our previous work, the exomoon is likely to be too small to observe directly, but its gravitational effects on the host planet will be quite visible.  Consider placing a moon around the planet such that the planet-moon orbital plane is the same as the star-planet orbital plane.  Due to the moon's presence, the planet will exhibit reflex motion, or put simply a wobble, in this plane.  From figure 1, it is clear that a component of this perturbation will be in the $z$-direction.

The motion along the $z$-axis is of particular interest because we have already discussed how sensitive the transit duration is to the distance $q$.  Any motion in this direction will cause $q$ to get periodically larger and smaller.  Ergo, the transit duration will vary.

In the proposed case, we consider an exomoon with an orbit coaligned to a planet-star plane at an orbital inclination angle $i$, where $i \leq 90^{\circ}$.  It is clear to see that there will, in general, always be a component  of $z$-axis wobbling motion for other moon inclinations.  One of the special cases where this will not occur is when $i=90^{\circ}$ and the moon takes a coplanar orbit, which is precisely the case we considered in our previous work.

\section{The Updated Model}
\subsection{Basic setup}

In our updated model, we consider a planetary orbit where $i \leq 90^{\circ}$.  We derive the total TDV effect by considering several stages of geometric manipulation of the planet's motion.  We use the same reference axes in figure 1 where the observer is at $y = +\infty$, the planet-moon barycentre's true anomaly is $f_P$, and the planet's true anomaly around the planet-moon barycentre is $f_W$.  In appendix A, we briefly consider the effects of inclined moon orbits, but for the mathematically simpler (and possibly more probable) case of a co-aligned moon orbit, it is shown that we may write the position of the planet as:

\begin{align}
x(f_P,f_W) &=  r_P \cos(f_P + \varpi_P) + r_W \cos (f_W+\varpi_W) \nonumber \\
z(f_P,f_W) &= [r_P \sin(f_P+\varpi_P) + r_W \sin (f_W + \varpi_W)] \cos i
\end{align}

Where $r_P$ and $r_W$ are the planet to star and planet to planet-moon barycentre separations respectively.  Note that we adopt the same notation as our previous paper\footnote{Except for tilting the orbital plane into the +$z$ direction rather than the negative}.  $q$ is nominally given by $r_P \cos i$ (see equation 1), but here we consider that $q$ has been perturbed by the moon's presence to a new value given by $q'$.  We denote the perturbation itself by $\Delta q$.  It can be seen that $q' = z(f_P \rightarrow f_{mid},f_W) $, where $f_{mid}$ is the true anomaly at the moment of mid-transit, given by $f_{mid} = \pi/2 - \varpi_P$.

Without a moon, we would simply have $z(f_P) = r_P \cos i \sin(f_P + \varpi_P)$ and thus $z(f_P \rightarrow f_{mid}) = q = r_P \cos i$, i.e. no perturbation.  Defining $q' = q + \Delta q$, according to the directions in figure 1, we now have $\Delta q = q' - q$:

\begin{equation}
\Delta q = r_W \sin (f_W + \varpi_W) \cos i
\end{equation}

Note that $\Delta q > 0$ represents a shift in the $+z$-direction whereas $\Delta q < 0$ is a shift in the $-z$-direction.  As expected, for $i \rightarrow 90^{\circ}$, we have $\Delta q \rightarrow 0$.  For a highly inclined moon orbit, it is worth noting that the maximal value of $\Delta q$ will be $\Delta q = a_W$.

\subsection{Prograde versus retrograde orbital motion}

Consider the planet-moon barycentre moving in the $+x$-direction in figure 1.  For a prograde orbit, the velocity of the planet around the planet-moon barycentre must be in the $+x$-direction when it is at position (1).  At position (1), the transit impact parameter has increased and thus the transit duration has shortened.  At the same time, the planet's wobble velocity is additive to the planet-moon barycentre velocity around the host star, and so the transit duration is further shortened.  Thus for prograde orbits, it can be seen that the TIP- and V-components are additive. The opposite is true for retrograde orbits.

\subsection{Derived total TDV effect}

In appendix C, we evaluate the total TDV effect in the case of $e_S, \alpha, \beta, \gamma = 0$ and $0 \leq e_P <1$.  We are able to show that the r.m.s. amplitude of the TDV signal is given by:

\begin{equation}
\delta_{TDV} \simeq \Big[\underbrace{\frac{a_W a_P \cos^2i}{(R_*+R_P)^2-a_P^2 \cos^2 i}}_\text{TIP-component} \pm \underbrace{\frac{2 \pi a_W}{P_S} \frac{1}{v_{B\bot}}}_\text{V-component}\Big] \cdot \frac{\bar{\tau}}{\sqrt{2}} 
\end{equation}

Where $a_W$ is the semi-major axis of the planet's orbit around the planet-moon barycentre, $a_P$ is the semi-major axis of the planet-moon barycentre's orbit around the host star, $i$ is the orbital inclination angle of the planet-moon barycentre, $v_{B\bot}$ is the projected velocity of the planet-moon barycentre across the face of the star during transit, $P_S$ is the orbital period of the satellite and $\bar{\tau}$ is transit duration of the planet in the absence of a moon.

The positive sign refers to prograde moon orbits and the negative signs refers to retrograde orbits.  It can be seen that the TDV effect has two dominant terms.  The first term is the TIP-component is $\propto a_S M_S$, which is the same as TTV's proportionality (where $M_S$ is the mass of the satellite).  The second term is the V-component and is $\propto M_S a_S^{-1/2}$, as found in our previous paper.  A summary of the properties of the three known transit timing effects due to an exomoon can be seen in table 1.

\begin{table}
\caption{\emph{Summary of key properties of the three known transit timing effects due to an exomoon.}} % title of Table
\centering % used for centering table
\begin{tabular}{c c c c} % centered columns (4 columns)
\hline\hline %inserts double horizontal lines
& TTV & TDV-V & TDV-TIP \\ [0.5ex] % inserts table
%heading
\hline % inserts single horizontal line
Type of effect & Positional & Velocity & Positional \\
Direction & $\hat{x}$ & $\hat{x}$ & $\hat{z}$ \\
Proportionality & $M_S a_S$ & $M_S a_S^{-1/2}$ & $M_S a_S$ \\
Relative phase & $0$ & $\pi/2$ & $\pm \pi/2$ \\ [1ex]
\hline\hline %inserts single line
\end{tabular}
\label{table:nonlin} % is used to refer this table in the text
\end{table}

\begin{table*}
\caption{\emph{Predicted TTV and TDV (both V- \& TIP- components) rms amplitudes due to a $1 M_{\bigoplus}$ exomoon at 1/3 the Hill radius, for a selection of the best candidate transiting planets.  System parameters are taken from various references, which are shown.}} % title of Table
\centering % used for centering table
\begin{tabular}{c c c c c} % centered columns (4 columns)
\hline\hline %inserts double horizontal lines
Planet & $\delta_{TTV}$/s & V-part of $\delta_{TDV}$/s & TIP-part of $\delta_{TDV}$/s & Reference \\ [0.5ex] % inserts table
%heading
\hline % inserts single horizontal line
HAT-P-11b & 19.19 & 22.54 & 0.40 & \citet{bak09} \\
GJ436b & 14.12 & 13.68 & 1.30 &\citet{alo08} \& \citet{tor07} \\
CoRoT-Exo-4b & 7.58 & 9.15 & 0.00 & \citet{aig08} \\
OGLE-TR-111b & 4.63 & 7.32 & 0.11 & \citet{dia08} \\
HAT-P-1b & 4.58 & 6.82 & 0.47 & \citet{joh08} \\
HD149026b & 3.61 & 9.76 & 0.00 & \citet{win07} \\
Lupus-TR-3b & 3.28 & 5.19 & 0.07 & \citet{wel08} \\
WASP-7b & 3.26 & 5.88 & 0.00 & \citet{hel09} \\
HD17156b & 3.07 & 1.06 & 0.43 & \citet{bar07} \\
TrES-1b & 3.04 & 5.95 & 0.05 & \citet{wina07} \\
HD209458b & 2.97 & 5.95 & 0.07 & \citet{kip08} \\
XO-5b & 2.65 & 4.69 & 0.17 & \citet{bur08} \\
HAT-P-4b & 2.54 & 8.34 & 0.00 & \citet{kov07} \\
HD189733b & 1.52 & 2.96 & 0.16 & \citet{winb07} \& \citet{bea08} \\ 
XO-3b & 0.41 & 0.87 & 0.07 & \citet{win08} \\ [1ex]
\hline\hline %inserts single line
\end{tabular}
\label{table:nonlin} % is used to refer this table in the text
\end{table*}

\section{Implications}
\subsection{TDV's inclination dependence}

By considering orbital inclination and the transit impact parameter, we have demonstrated that an additional TDV effect exists, which we have labelled as the TIP-component.  It may be tempting to assume that inclined orbits therefore improve the TDV effect due to an exomoon (assuming a prograde orbit) and perhaps the best exoplanets candidates for detection purposes would be near-grazing transits.  However, we counter that naive supposition by pointing out that the transit duration itself decreases with higher impact parameters and it can be seen in equation (4) that both the V- and TIP-components have dependancies on $\bar{\tau}(b)$.

We may rightly ask whether there is a certain value of impact parameter which enhances the TDV signal optimally.  By differentiating equation (4) with respect to $b$ and making some approximations, we find that the optimal value of $b$ occurs for $b \simeq (1+k)$, where $k$ is the ratio-of-radii.  In other words, the TDV signal is enhanced for partial transits where the impact parameter is so high the lightcurve takes a V-shape.

In figure 2, we plot the variation of the TDV effect for a prograde $1 M_{\oplus}$ exomoon around a hypothetical planet.  We use the same hypothetical planet as in our previous, K09, i.e. an identical system to GJ436b except the orbital period is 35.7 days and the eccentricity is zero.  As seen in figure 2, the TDV effect slowly drops off for increasing $b$ until we reach the partial-transit regime where the TDV effect becomes extremely large due to the TIP-component dominating.

However, the probability of detecting a transit which is only partially transiting is very small due to two reasons: 1) geometrically the inclination range is very small 2) V-shaped lightcurves are usually rejected as a planetary candidate and labelled as a grazing eclipsing binary.  For these reasons, we consider the optimal planets for detecting an exomoon to have the longest transit duration possible, i.e. $b \simeq 0$.

\begin{figure}
\begin{center}
\includegraphics[width=8.4 cm]{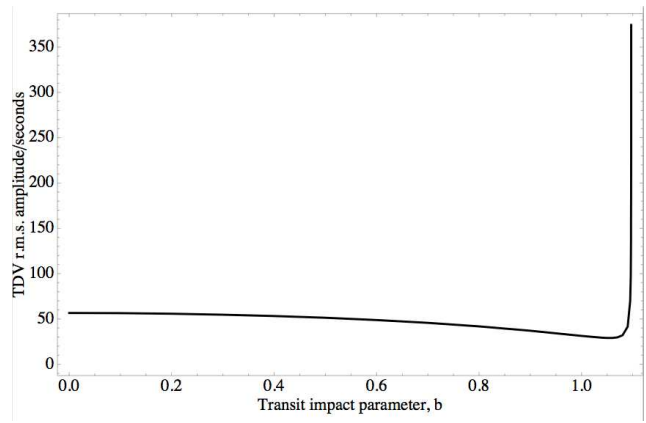}
\caption{\emph{For a hypothetical planet-moon system, we plot the variation of the TDV effect with transit impact parameter.  Whilst partially transiting planets seem to offer enhanced signals, they are also highly unlikely to be detected and thus the case of $b=0$ is more likely to be the optimal condition.}} \label{fig:fig2}
\end{center}
\end{figure}

\subsection{Determining $P_S$ and $M_S$}

What are the consequences for determing both $a_S$ and $M_S$ separately, as proposed in our last paper?  Certainly taking the simple ratio of the TDV and TTV effect will not provide precisely the same equation as we predicted before.  From \citet{kip09}, the TTV r.m.s. amplitude, for a circular moon orbit, is given by:

\begin{equation}
\delta_{TTV} = \frac{1}{\sqrt{2}} \frac{a_S M_S \sqrt{a_P}}{M_{PRV} \sqrt{G (M_* + M_{PRV})}} \frac{1}{\Upsilon}
\end{equation}

Where $\Upsilon$ accounts for the planet's orbital eccentricity.  Taking the ratio of TDV to TTV, we have:

\begin{align}
\eta &= \frac{\delta_{TDV}}{\delta_{TTV}} \nonumber \\
\eta &= \underbrace{\frac{\sqrt{G a_P (M_* + M_{PRV})} \Upsilon \bar{\tau} \cos^2i}{(R_* + R_P)^2-a_P^2 \cos^2i}}_\text{`constant'} \pm \underbrace{\frac{2 \pi \bar{\tau}}{P_S}}_\text{`info'}
\end{align}

The TIP-component makes $\eta$ change from simply being inversely proportional to $P_S$ to being inversely proportional to $P_S$ plus a constant.   We highlight the following key points about this equation:

\begin{itemize}
\item[{\tiny$\blacksquare$}] The `constant' is not a function of the moon's properties whatsoever, i.e. it depends on the planetary properties only.
\item[{\tiny$\blacksquare$}] In the case of $i = 90^{\circ}$, the `constant' $\rightarrow 0$, retrieving the original equation for $\eta$, as derived in K09.
\item[{\tiny$\blacksquare$}] The `constant' may be calculated independently of the moon's properties in a reliable way and thus $P_S$ may still be calculated, providing we assume $e_S = 0$ and $\alpha$, $\beta$, $\gamma = 0$.
\end{itemize}

Equation (6) tells us that it is still possible to evaluate $P_S$ by simply taking the ratio of the TDV and TTV effects, provided we make certain assumptions about the moon's orbit.  The `constant' term quoted above is not a function of the moon's properties, but in reality it is actually a very weak function since we have made the assumption $a_P \gg a_W$, which is certainly a very valid assumption to make.

Although a large TIP-component can enhance the detectable signature of an exomoon, if the TIP-component is greatly larger than the V-component, then our ability to accurately estimate $P_S$ and hence $M_S$ will diminish.  Ideally, the V-component should remain the dominant term for accurate determination of these parameters.

For a system of interest like GJ436b, the constant is $\sim 0.1$ suggesting that TTV is an order of magnitude stronger than the TIP-component of the TDV effect.  Although this is small, the key point is that it is additive and can be larger than the V-component in certain special cases, notably grazing transits.

We conclude that our previous statement that the ratio of TDV and TTV can be used to determine the moon's orbital distance and mass still remains true, provided we assume a circular co-aligned exomoon orbit.  For systems with exomoon eccentricity and inclinations, there will be insufficient information through timing alone to solve for all of these parameters.

\subsection{Determining the sense of an exomoon's orbital motion}

Observations of transit timing may also permit the determination of whether an exomoon is in a prograde or retrograde orbit, given sufficient signal-to-noise.  We will illustrate this possibility by referring to a hypothetical example of an exomoon detection.  We use the same example as in our last paper, K09, where we considered shifting GJ436b to an circular orbit of 35.7 days period and add a $1 M_{\oplus}$ exomoon on a $2.5$ day orbit.  We use the same impact parameter as measured for GJ436b by \citet{tor07}, i.e. $b = 0.848$.

Consider we measure the TTV r.m.s. amplitude of this planet to be $137.4 \pm 0.4$ seconds and the total r.m.s. TDV amplitude would be $39.7 \pm 0.8$ seconds giving $\eta = 0.289 \pm 0.020$.  Note that we have assumed the duration is measured to half the precision of the mid-transit time and assumed timing errors in-line with the capabilities of forthcoming missions.  Based on the known inclination and planetary properties, we are able to evaluate the $\eta$ `constant' term to be $0.02739$ to a negligible error (since this can be refined by compositing multiple transits).  Therefore the `info' component of $\eta$ is either $0.261 \pm 0.020$, if prograde, or $0.316 \pm 0.020$, if retrograde.  This corresponds to an exomoon period of either $2.52 \pm 0.14$ days or $2.08 \pm 0.14$ days for prograde and retrograde respectively, differing by 3-sigmas.  With the TTV measured to a signal-to-noise of over 350, it would not be difficult to use multiple TTV measurements to identify which of these periods is permitted by the frequency of the data points, which in this case is the prograde orbit.

It is important to remember that this calculation was done for a planet-moon orbital plane which is coaligned to the star-planet orbital plane.  Slight moon inclinations of $\lesssim 5^{\circ}$ would not change the result significantly but very large inclinations would severely disrupt this calculation's accuracy.  We propound that exomoons of low inclination angles would be identifiable by a planet-moon eclipse which should be observable in the lightcurve, as predicted by \citet{cab07}.

We therefore propose that it should be possible for future observations to not only detect an exomoon and determine its mass, but also provide a confident deduction of the sense of orbital motion.  Although this determination will likely require photometry at the limit of planned missions, it seems likely that once an exomoon is detected a more in-depth investigation would be able to answer the question of orbital sense of motion conclusively.

\section{Discussion and Conclusions}

We have shown that an exomoon around a transiting exoplanet should induce a transit duration variation effect with two dominant components.  One of these components is due to the moon altering the velocity of the host planet, which we label as the V-component.  The second constituent is due to the impact parameter of the transiting planet varying as a result of the moon's presence, which we label as the TIP-component.

In table 2, we have evaluated the V- and TIP- components for a list of targets as in our previous paper.  The table suggests that the TIP-component is often an order of magnitude less than the V-component, but can therefore we several seconds for some targets.  We do not anticipate this additional component to be a hurdle in determining the moon's mass and orbital distance since the dominant effect on the $\eta$ parameter is to introduce an additive constant, which is independent of the moon's properties.

The TDV effect can be markedly increased for prograde moons which improves their detectability.  For grazing transits with highly inclined moons, the planet could even go through epochs where it no longer transits at all, although we do not expect this to be a typical situation.  Furthermore, we predict that including the TIP-component may allow for the determination of the orbital sense of motion of an exomoon around an exoplanet.  We do however stress that such a determination would require very precise, but feasible, photometry.  We believe this paper further demonstrates the feasibility of detecting an exomoon, but outlines the great care and understanding required to complete the analysis.

\section*{Acknowledgments}

DMK is supported by STFC, University College London and HOLMES ANR-06-BLAN-0416. The author would like to thank Jean Schneider for technical discussions on the subject of transit timing effects.  Special thanks to the reviewer of this paper whose comments have greatly improved the manuscript.

\appendix

\section{Mathematical Treatment}

In appendix A, we will derive the two components of the TDV effect separately.  The two effects will later be combined in appendix B.  The total TDV effect will then be evaluated in the case of a circular, coaligned moon in appendix C.  The error regarding one of our key assumptions will be calculated in appendix D.

\subsection{Planetary wobble motion}

From figure 1, it is clear that the change we care about, $\Delta q$ is in the $+z$-direction.  In order to create a generally oriented orbital plane, we do so in several steps.  In our first step, we consider the planet-moon centre of mass frame and employ the same geometric model as that of our previous paper where we consider an elliptical orbit with a centre at the origin of an $x'$-$y'$-$z'$ co-ordinate system, the $S'$ frame.  We may write the position of the planet, as a function of its true anomaly within this frame ($f_W$), as:

\begin{align}
x' &= a_W e_W + r_W \cos f_W \nonumber \\
y' &= r_W \sin f_W \nonumber \\
z' &= 0
\end{align}

where
\begin{equation}
r_W = \frac{a_W (1-e_W^2)}{1+e_W \cos f_W}
\end{equation}

In the same manner described by \citet{kip08}, we then transform these positional coordinates with i) a counter-clockwise rotation about $z'$-axis by the position of pericentre angle, $\varpi_W$ (the $S_1'$ frame) ii) a translation of the planet-moon barycentre to the origin (the $S_2'$ frame).  These two transformations give us:

\begin{align}
x_2' &= r_W \cos(f_W + \varpi_W) \nonumber \\
y_2' &= r_W \sin(f_W + \varpi_W) \nonumber \\
z_2' &= 0
\end{align}

It is worth noting that in the case of $e_W \simeq 0$, which we would anticipate to be the typical scenario, we have $x_2' = a_W \cos f_W$ and $y_2' = a_W \sin f_W$.

In this system, we allow the orbital arrangement of the moon to be rotated by the three Euler angles, $\alpha$, $\beta$ and $\gamma$ with respect to the star-planet plane.  We choose to employ the conventional $z$-$x$-$Z$ convention for the Euler system.  The Euler rotation angles are ideal because any general rotation in three dimensions can always be written in terms of these three angles.  

We denote the ${x_2',y_2',z_2'}$ position of the planet away from the planet-moon barycentre origin by the vector $\mathbf{X}_2'$, then after the Euler rotations we have a new positional vector given by:

\begin{equation}
\mathbf{X}_3' = \mathbf{R}_{\textrm{Euler}}(\alpha,\beta,\gamma) \mathbf{X}_2'
\end{equation}

where $\mathbf{R}_{\textrm{Euler}}$ is the Euler rotation matrix.  Since $z_2' = 0$, we may write:

\begin{align}
x_3' &= [c_{\alpha} c_{\gamma}  - s_{\alpha} c_{\beta} s_{\gamma}] \cdot x_2' + [-c_{\alpha} s_{\gamma} - s_{\alpha} c_{\beta} c_{\gamma}] \cdot y_2' \nonumber \\
y_3' &= [s_{\alpha} c_{\gamma} + c_{\alpha} c_{\beta} s_{\gamma}] \cdot x_2' + [-s_{\alpha} s_{\gamma} + c_{\alpha} c_{\beta} c_{\gamma}] \cdot y_2' \nonumber \\
z_3' &= s_\beta s_\gamma \cdot x_2' + s_\beta c_\gamma \cdot y_2'
\end{align}

\begin{figure}
\begin{center}
\includegraphics[width=8.4 cm]{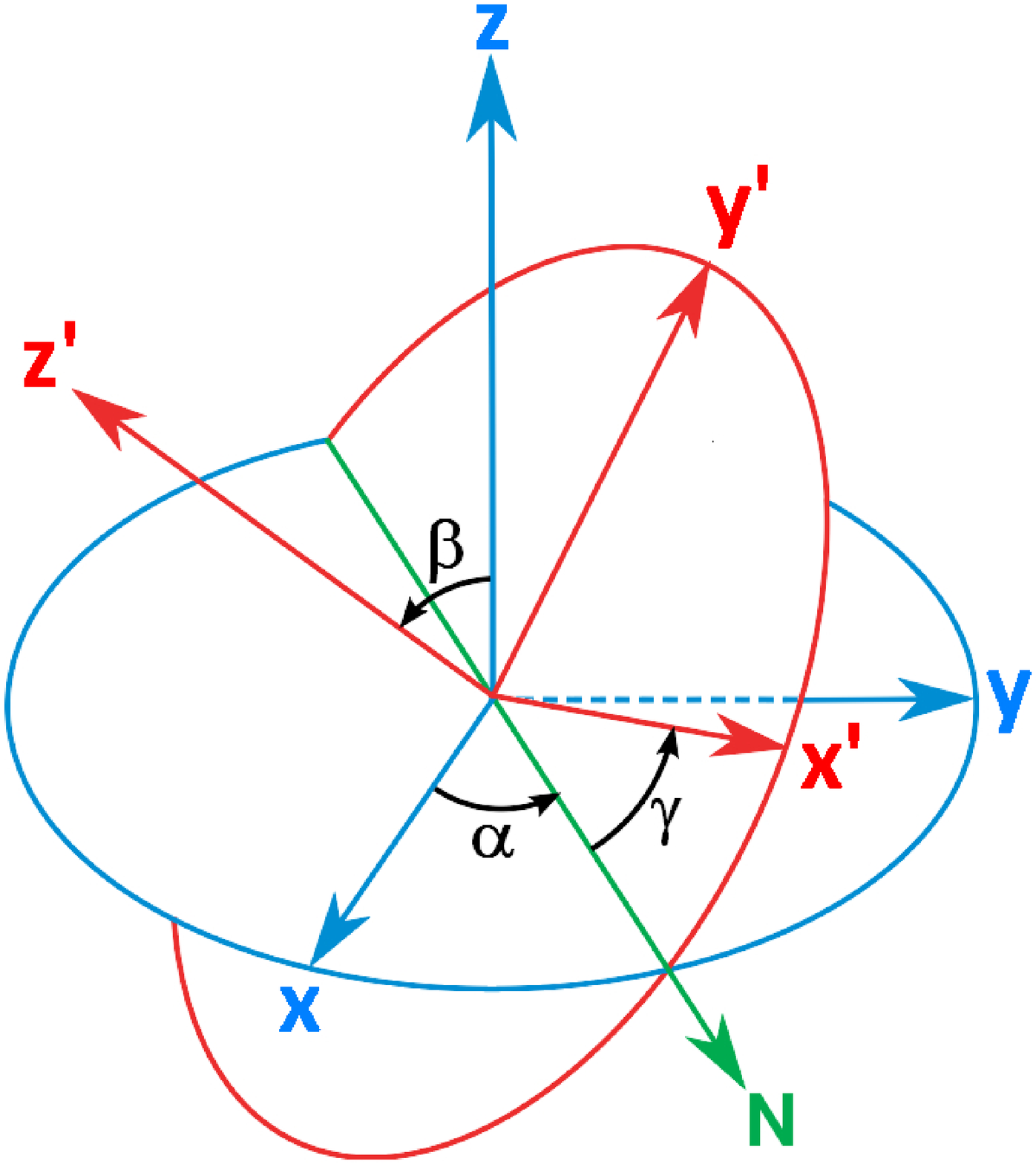}
\caption{\emph{Euler angles of our rotational scheme.}} \label{fig:fig3}
\end{center}
\end{figure}

We have therefore found the position of the planet away from the planet-moon barycentre as a function of $f_W$.

\subsection{Planet-moon barycentric motion}

Consider the orbit of a point mass around the host star.  We denote this point mass by the coordinates ${\tilde{x},\tilde{y},\tilde{z}}$ (the $\tilde{S}$ frame).  Placing the origin at the centre of the ellipse, we can write:

\begin{align}
\tilde{x} &= a_P e_P + r_P \cos f_P \nonumber \\
\tilde{y} &= r_P \sin f_P \nonumber \\
\tilde{z} &= 0
\end{align}

We then rotate for position of pericentre and translate so that the star is at the origin.

\begin{align}
\tilde{x_2} &= r_P \cos(f_P+\varpi_P) \nonumber \\
\tilde{y_2} &= r_P \sin(f_P +\varpi_P) \nonumber \\
\tilde{z_2} &= 0
\end{align}

For the moment we will not include the orbital inclination angle, but return to it later.

\subsection{Overall motion of planet}

Earlier on we defined the position of the planet, in relation to the planet-moon barycentre as ${x_3',y_3',z_3'}$.  So the true position of the planet is the vector for the barycentre's position, $\tilde{\mathbf{X}_2}$ added to the vector for the relative position of the planet, with respect to the barycentre, $\mathbf{X}_3'$.  This gives us $\mathbf{X}_{A} = \tilde{\mathbf{X}_2}+\mathbf{X}_3'$.

\begin{align}
x_A &= \tilde{x_2} + x_3' \nonumber \\
y_A &= \tilde{y_2} + y_3' \nonumber \\
z_A &= \tilde{z_2} + z_3'
\end{align}

Finally, we now rotate about the $x_A$-axis in a counter-clockwise sense by an angle $(\pi/2 -i)$, in order to be consistent with our defined system shown in figure 1.  Note that in our previous paper we performed this rotation in the clockwise direction and we choose to do the opposite direction here for mathematical simplicity.  Of course, the choice of direction does not affect the final result.

\begin{align}
x &= x_A \nonumber \\
y &= y_A \sin i - z_A \cos i \nonumber \\
z &= y_A \cos i + z_A \sin i
\end{align}

\subsection{Application to a co-aligned moon orbit}

Although the Euler angles can play a significant role in enhancing the TDV effects, we predict that most exomoons will not exhibit significant inclination deviations from the planet-star plane and so we decide to set all of these angles to zero.  This is also valid because it would be impossible to determine all of the Euler angles from just two measurements in any case.  Thus we have:

\begin{align}
x &= \tilde{x_2} + x_2' \nonumber \\
y &= [\tilde{y_2} + y_2'] \sin i \nonumber \\
z &= [\tilde{y_2} + y_2'] \cos i
\end{align}

For an observer at infinite $+y$, the $y$-component is never seen.  So we have:

\begin{align}
x(f_P,f_W) &=  r_P \cos(f_P + \varpi_P) + r_W \cos (f_W+\varpi_W) \nonumber \\
z(f_P,f_W) &= [r_P \sin(f_P+\varpi_P) + r_W \sin (f_W + \varpi_W)] \cos i
\end{align}

From figure 1, $q' = z(f_P \rightarrow f_{mid},f_W)$ and $\Delta q = q' - q$ where $q$ is the offset in the $z$-direction when no moon is present.  Removing the perturbation terms, we can derive the moonless quantity to be $q = r_P \sin(f_{mid}+\varpi_P) \cos i = r_P \cos i$.  The two equations therefore give the perturbation magnitude to be:

\begin{equation}
\Delta q = r_W \sin (f_W + \varpi_W) \cos i
\end{equation}

\subsection{TIP-component of the TDV effect}

Here, we will derive approximate equations for the TIP-component of the TDV effect.  For simplicity, we may write down the case of $e_W = 0$.  We also assume $\alpha = \beta = \gamma = 0$, but $i \neq 90^{\circ}$.  Ergo, this is not the same case considered in our previous research where $i=90^{\circ}$.  Although we could write the down the expressions for the most general case, to do seems pointless given that there will not be enough information to solve for all of these parameters in any case.  For the stated assumptions:

\begin{align}
\Delta q(f_W) = a_W \cos i \sin f_W \nonumber \\
v_{W\bot}(f_W) = \pm |v_{W\bot}| \sin f_W
\end{align}

\emph{where the $\pm$ symbol refers to prograde/retrograde exomoon orbits respectively}

Although in reality the magnitude of the vector connecting the planet to the star's centre is no longer $a_P$, the maximum by which it can be modified will be given by $a_W$.  Since $a_P \gg a_W$, we will assume that this changing magnitude vector does not significantly affect the TDV amplitude relative to the change in $q$.

As a result of changing $q$, the observed transit impact parameter appears to change.  Without a moon present, the impact parameter should be simply given by:

\begin{equation}
b = \frac{q}{R_*}
\end{equation}

But now the altered effective impact parameter will be given by:

\begin{equation}
b'(f_W) = \frac{q'}{R_*} = \frac{q+\Delta q(f_W)}{R_*}
\end{equation}

We now consider that the TDV signal due to the TIP effect is given by the observed duration ($\tau$) minus the expected duration:

\begin{equation}
\textrm{TDV}(f_W) = \tau(b') - \tau(b) = \tau(b') - \bar{\tau}
\end{equation}

\subsection{Modification to the V-component}

\citet{kip09} proposed that the velocity of a planet during a transit is modified by the presence of an exomoon.  We label this TDV effect as the V-component.  This effect is in addition to the previously detailed TIP-component.  We will now derive the modification to the V-component in the presence of orbital inclinations.  In our previous paper, we defined:

\begin{align}
\textrm{TDV}(f_W) &= \tau(f_W) - \bar{\tau} \nonumber \\
\lim_{i \rightarrow \pi/2} \textrm{TDV}(f_W) &= \Big(\frac{v_{B\bot}}{v_{B\bot} + v_{W\bot}(f_W)} - 1\Big) \cdot \bar{\tau}
\end{align}

In our orbital setup, we assume the planet is moving in the positive $x$-direction.  Therefore, we need to know what component of $v_W$ remains in this direction given a rotation through the three Euler angles.  Taking the original vector $v_W = \{v_{W\bot},0,0\}$ and rotating we find that the $x$-component is modified to:

\begin{align}
v_{W\bot} &\rightarrow [\cos(\alpha) \cos(\gamma)  - \sin(\alpha) \cos(\beta) \sin(\gamma)] v_{W\bot} \nonumber \\
v_{W\bot} &\rightarrow \varphi(\alpha,\beta,\gamma) \cdot v_{W\bot}
\end{align}

This modifies our velocity TDV component to:

\begin{equation}
\lim_{i \rightarrow \pi/2} \textrm{TDV}(f_W) = \Big(\frac{v_{B\bot}}{v_{B\bot} + \varphi(\alpha,\beta,\gamma) \cdot v_{W\bot}(t)} - 1\Big) \cdot \bar{\tau}
\end{equation}

But note that this does not include the TIP-component.  Now that we have written down the TDV signal for both the V- and TIP-components independently, the next step is to combine the two.

\section{The Total TDV effect}

We now consider the total effect.  The V-component is just a factor which modifies the duration, so this can be applied after the TIP-component.  By employing this ordering, we are able to write the transit duration, in the the general case, as:

\begin{equation}
\tau(f_W) = \Big(\frac{v_{B\bot}}{v_{B\bot} + \varphi(\alpha,\beta,\gamma) \cdot v_{W\bot}(f_W)} \Big) \cdot \tau(b')
\end{equation}

It would useful at this point to have $\tau(b')$ written as some factor multiplied by the mean transit duration $\bar{\tau}$, where it is understood that $b'(f_W)$.  Let:

\begin{equation}
\varepsilon(b') = \frac{\tau(b')}{\bar{\tau}}
\end{equation}

We note that $\tau$ gets larger as $b$ approaches zero.  This occurs for $\Delta q$ being positive.  Thus, $\varepsilon > 1$ for $\Delta q > 0$ and vice versa.  Our TDV may now be written as:

\begin{equation}
\textrm{TDV}(f_W) = \Big(\frac{\varepsilon(b') \cdot v_{B\bot}}{v_{B\bot} + \varphi(\alpha,\beta,\gamma) \cdot v_{W\bot}(f_W)} - 1\Big) \cdot \bar{\tau}
\end{equation}

Appreciating that $v_{W\bot} \ll v_{P\bot}$, this expression may be approximated to:

\begin{equation}
\textrm{TDV}(f_W) \simeq \Big[(\varepsilon-1) - \frac{\varepsilon \cdot v_{W\bot}}{v_{B\bot}}\Big] \cdot \bar{\tau}
\end{equation}

We can now see that the TDV signal has two clear components.  In the absence of any TIP-component, $\varepsilon \rightarrow 1$ and hence we recover the original TDV effect predicted in our previous paper.  Since $\varepsilon$ is a value close to unity, we choose to write it as:

\begin{equation}
\varepsilon = 1 + \varrho
\end{equation}

Where it is understood that $\varrho$ is small compared to unity.  This now gives us:

\begin{align}
\textrm{TDV}(f_W) &\simeq \Big[\varrho - \frac{v_{W\bot}}{v_{B\bot}} - \varrho \cdot \frac{v_{W\bot}}{v_{B\bot}}\Big] \cdot \bar{\tau} \nonumber \\
\textrm{TDV}(f_W) &\simeq \Big[\varrho(f_W) - \frac{v_{W\bot}(f_W)}{v_{B\bot}}\Big] \cdot \bar{\tau}
\end{align}

Consider the case of $\Delta q > 0$, as mentioned earlier this means $\varrho > 0$ and so the first term is positive.  $\Delta q$ is positive when $v_W(f_W)$ goes negative for prograde orbits.  So the second term must be negative too, and thus we have a double negative which equals a positive.  Thus we confirm that when for prograde orbits the TDV effect is additive.  

It is also clear that the total effect is dominated by two terms which are linearly additive.  These means the TIP-component can significantly increase TDV signals due to prograde exomoons.  On the other hand, it can significantly dampen any effect for retrograde moons.  Infact, if the TIP-component is very large for a retrograde, it may change the sign of the effect completely meaning that instead of TDV lagging behind TTV by $\pi/2$, the opposite is true.

\section{Evaluation of the TDV effect}
\subsection{Case of $e_S,\alpha, \beta, \gamma = 0$ \& $0 \leq e_P < 1$}

In order to ascertain the proportionality and magnitude of the total TDV effect, let us first assume the transit duration is given by the circular equations described by \citet{sea03}.  Using equation (16) from this paper and replacing $a_P \cos i$ with $q$:

\begin{equation}
\tau \simeq \frac{P_P R_*}{\pi a_P} \sqrt{\Big(1 + \frac{R_P}{R_*}\Big)^2 - \Big(\frac{q}{R_*}\Big)^2}
\end{equation}

Which gives us:

\begin{equation}
\varepsilon(b')^2 = \frac{\Big(1 + \frac{R_P}{R_*}\Big)^2 - \Big(\frac{q+\Delta q}{R_*}\Big)^2}{\Big(1 + \frac{R_P}{R_*}\Big)^2 - \Big(\frac{q}{R_*}\Big)^2}
\end{equation}

Even if we use the equations for the approximate transit duration due to an eccentric orbit as presented by \citet{for07} by their equation (1), the same value of $\varepsilon^2$ is derived.  Thus the following derivation holds true for planets on eccentric orbits.  Expanding out the brackets we have:

\begin{equation}
\varepsilon(b')^2 = \frac{\Big(1 + \frac{R_P}{R_*}\Big)^2 - \Big(\frac{q}{R_*}\Big)^2 - 2 \Big(\frac{q \Delta q}{R_*^2}\Big) - \mathcal{O}\Big(\frac{\Delta q}{R_*}\Big)^2}{\Big(1 + \frac{R_P}{R_*}\Big)^2 - \Big(\frac{q}{R_*}\Big)^2}
\end{equation}

Assuming $\mathcal{O}\Big(\frac{\Delta q}{R_*}\Big)^2$ is small and $a_P \gg a_W$, we may write:

\begin{equation}
\varepsilon(b') \simeq \sqrt{1 - \frac{2 a_W a_P \cos^2i \sin f_W}{(R_*+R_P)^2-a_P^2 \cos^2 i}}
\end{equation}

Unfortunately, evaluating the integral of $\textrm{TDV}^2$ is non-trivial, even with these approximations.  Without a direct computation of this integral, we cannot evaluate the r.m.s. amplitude directly.  In order to estimate the r.m.s. amplitude, we assume the signal takes a sinusoidal form and thus the r.m.s. amplitude will be given by the normal amplitude divided by $\sqrt{2}$.  This approximation is particularly valid because we have assumed $e_S = 0$.  The maximum value of $\varepsilon$ and $v_{W\bot}$ is for $f_W = -\pi/2$.  We also take advantage of the fact we know $\varepsilon$ is close to unity and therefore we approximate the square root to find $\varrho_{max}$.

\begin{align}
\varepsilon_{max} &\simeq 1 + \frac{a_W a_P \cos^2i}{(R_*+R_P)^2-a_P^2 \cos^2 i} = 1 + \varrho_{max} \nonumber \\
\varrho_{max} &= \frac{a_W a_P \cos^2i}{(R_*+R_P)^2-a_P^2 \cos^2 i}
\end{align}

This gives us a max signal of:

\begin{equation}
\textrm{TDV}_{max} \simeq \Big[\frac{a_W a_P \cos^2i}{(R_*+R_P)^2-a_P^2 \cos^2 i} \pm \frac{2 \pi a_W}{P_S} \frac{1}{v_{B\bot}}\Big] \cdot \bar{\tau} 
\end{equation}

Or an r.m.s. signal of:

\begin{equation}
\delta_{TDV} \simeq \Big[\frac{a_W a_P \cos^2i}{(R_*+R_P)^2-a_P^2 \cos^2 i} \pm\frac{2 \pi a_W}{P_S} \frac{1}{v_{B\bot}}\Big] \cdot \frac{\bar{\tau}}{\sqrt{2}} 
\end{equation}

\section{Error on constant velocity assumption}

Throughout we have assumed that $v_{W\bot}$ does not change significantly over the course of the transit.  This essentially the same as assuming that $P_S \gg \bar{\tau}$.  For a normal circular orbit, the planetary transit would occur for $f_P$ changing by a quantity:

\begin{equation}
\Delta f_P = \arcsin\Big[\frac{(1+k)^2 + (a_P \cos i/R_*)^2}{\sin^2 i}\Big]
\end{equation}

If the moon and planet are on a circular orbit, and $P_S = \Xi P_P$, then the change in $f_W$ may be written as:

\begin{equation}
\Delta f_W = \frac{1}{\Xi} \cdot \arcsin\Big[\frac{(1+k)^2 + (a_P \cos i/R_*)^2}{\sin^2 i}\Big]
\end{equation}

The velocity of the planet wobble is given by:

\begin{equation}
v_{W \bot} = |v_{W\bot}| \sin f_W
\end{equation}

So the average velocity over the course of $\Delta f_W$ is given by:

\begin{align}
<v_{W\bot}>(f_{W,mid}) &= \frac{|v_{W\bot}|}{\Delta f_W} \int_{f_{W,mid} - 0.5 \Delta f_W}^{f_{W,mid} + 0.5 \Delta f_W} \sin f_W \textrm{ d}f_W \nonumber \\
& = |v_{W\bot}| \sin (f_{W,mid}) \Big[\frac{\sin (\Delta f_W/2)}{\Delta f_W/2}\Big]
\end{align}

In contrast, previously we assume a fixed constant velocity of $v_{W\bot} = |v_{W\bot}| \sin f_{W,mid}$.  So it seems we very slightly overestimate the average velocity during the transit, as expected.  The dominant error term can be written out by expanding the $\sin$ function:

\begin{equation}
v_{W\bot}(f_W) \simeq |v_{W\bot}| \sin (f_{W,mid}) - |v_{W\bot}| \sin (f_{W,mid}) \frac{\Delta f_W ^3}{48}
\end{equation}

Thus the fractional error in $v_{W\bot}$ is given by $(\Delta f_W ^3/48)$.  The error should be largest for close-in orbits and even for a 3-day hot-Jupiter with typical parameters we would expect this error to be less than 1 part in $10^3$.

\bsp

\label{lastpage}

\end{document}